\documentclass[aps,pre,preprint,superscriptaddress]{revtex4-1}  
\usepackage{graphicx}
\usepackage{subcaption}
\usepackage{amssymb}
\usepackage{amsmath}
\usepackage{newtxtext,newtxmath,lipsum}
\usepackage[shortlabels]{enumitem}

\begin{document}
	\title{Instability of single- and double-periodic waves in the fourth-order nonlinear Schr\"odinger equation}
	\author{N. Sinthuja} 
	\affiliation{Department of Nonlinear Dynamics, Bharathidasan University, Tiruchirappalli 620024, Tamil Nadu, India}
	
	\author{S. Rajasekar}
	\affiliation{Department of Physics, Bharathidasan University, Tiruchirappalli 620024, Tamil Nadu, India}
	
	\author{M. Senthilvelan}
	\affiliation{Department of Nonlinear Dynamics, Bharathidasan University, Tiruchirappalli 620024, Tamil Nadu, India}
	\email{velan@cnld.bdu.ac.in}
	\begin{abstract}
		\par   We compute the instability rate for single- and double-periodic wave solutions of a fourth-order nonlinear Schr\"odinger equation.  The single- and double-periodic solutions of a fourth-order nonlinear Schr\"odinger equation are derived in terms of Jacobian elliptic functions such as $dn$, $cn$, and $sn$. From the spectral problem, we compute Lax and stability spectrum of single-periodic waves.  We then calculate the instability rate of single-periodic waves (periodic in the spatial variable). We also obtain the Lax and stability spectrum of double-periodic wave solutions for different values of the elliptic modulus parameter. We also highlight certain novel features exhibited by the considered system. We then compute instability rate for two families of double-periodic wave solutions of the considered equation for different values of the system parameter. Our results reveal that the instability growth rate is higher for the double-periodic waves due to the fourth-order dispersion parameter when compared to single-periodic waves.
	\end{abstract}
	
	\maketitle
	\section{Introduction}
	The nonlinear Schr\"odinger (NLS) equation is one among the completely integrable systems that describes the dynamics of deep ocean waves, propagation of pulse in optical fibers and several other phenomena from the nonlinear perspective \cite{wang}. It has been shown that rogue waves (RWs) \cite{jcmar,gelash,kmr,kns,dsa,sr} are rational solutions of the NLS equation. Nonlinear Schr\"odinger equation play a major role in understanding the appearance of complex RWs in the ocean. The rogue or freak waves that appear in the ocean are unpredictable, and when they appear, they come up three times larger than the ambient waves and produce catastrophic effects. Initially, efforts have been made to study the optical RWs (optical analogues of sea waves) on a constant background. The appearance of RW is related to the modulational instability (MI) of the wave background \cite{wen,zh}. MI represents the formation of a train of localized waves due to the breakup of carrier waves. This occurs due to the exponentially increasing amplitude of the wave as a consequence of the competition between dispersive effects and nonlinearity \cite{gelash2,bio,kra,kra1,hao,sul,tob,liu}. Recently, it has been realized that the mathematical studies on RWs on periodic wave background would be more useful in analyzing the dynamics of RWs on the surface of the ocean, where the surface wave appear periodic. Employing Darboux method and imposing periodic waves as the seed solution, RWs have been generated on a background of periodic wave for the NLS equation \cite{chen1,sinthu2}. 
	
	Since the dynamics of RWs depends on the MI characteristics \cite{choudhuri,wang1,biondini,grinevich,bilman}, studies have also been initiated to investigate the MI of the periodic waves \cite{dsa,chen2,xup}. The stability of periodic standing waves have been computed for the NLS equation through Floquet-Bloch decomposition method \cite{deco1}. Later, the periodicity in both space and time (double-periodic) has been taken into account. RWs on the top of this doubly periodic environment was also investigated for the NLS equation \cite{chen3}. Very recently, Pelinovsky studied the instability of the double-periodic waves of the NLS equation, numerically and showed that the instability rate of the standing periodic waves is greater than the waves that are periodic, both spatially and temporally. \cite{pelinovsky}.
	
	The NLS equation accurately simulates the propagation of longer light pulses since, in this model, only the lower order dispersive (second-order) term and a nonlinearity term were augmented. Subsequently, the fundamental NLS equation was expanded to include self-frequency shift, self-steepening and higher order dispersion terms upto fifth order. All these terms are included to study the dynamics of the propagation of femto/pico second LASER pulses in optical fibers and examine the higher-order nonlinear effects \cite{akmv1,akmv2,a11,a2,a3,a4,a41,a5,cyan,a8}. The applications of these extended NLS equations are not only restricted to water waves and optics but are applied in other branches of physics as well \cite{mal1,mal2}. Many of these higher-order NLS equations are integrable and different kinds of localized solutions have also been constructed for these equations. Further, it has been shown that the third- and fifth-order (odd order) operators in this family of equations give different features to the solutions, whereas even ordered operators do not. For more details one may refer the works \cite{zhj,chowdury}. Recently, RWs on the periodic \cite{a41,sinthu3} and double-periodic wave backgrounds \cite{sin} have been studied for a few of these higher-order NLS equations as well.  
	
	In this article, we investigate the fourth-order NLS equation and analyze the spectral stability problem and MI of this equation. Initially, we deduce the eigenvalues of single-periodic wave solutions (periodic in $x$) by solving the Lax pair equations. Then, we compute the values of certain unknown parameters that occur in the solution. Next, we add a linear perturbation term to the periodic wave solutions. With the help of this perturbation, we derive two linearized equations, whose solutions describe the linear stability/instability of the waves that are periodic in space. We then relate the variables of these equations with the spectral parameter. By suitably augmenting a second spectral parameter, we derive the spectral stability problem and numerically evaluate the rate of instability of $dn$- and $cn$- periodic waves. Our results show that when the elliptic modulus is varied from zero to one, the growth rate of instability of $dn$-periodic waves decreases and a similar observation is also observed for the $cn$-periodic waves. We then compute the instability of double-periodic wave solutions using Floquet theory. Our results confirm that the instability rate of single-periodic wave is higher when compared to double-periodic wave solutions.
	
	We present the work as follows: In Section 2, we present the Lax pair of a fourth-order NLS equation. In Section 3, we compute the instability rate for single-periodic waves. In Section 4, we calculate the instability rate for double-periodic waves. We present our conclusion in Section 5.
	
	\section{Model and Lax pair}
	Let us consider a fourth-order NLS equation,
	\begin{align}
		iq_t+\frac{1}{2}q_{xx}+|q|^2q+\gamma(q_{xxxx}&+6\bar{q}q^2_x+4|q_x|^2q+8|q|^2q_{xx}
		\nonumber\\&+2q^2\bar{q}_{xx}+6|q|^4q)=0,
		\label{s21}
	\end{align}
	in which $\bar{q}=\bar{q}(x,t)$ is the complex conjugation of $q=q(x,t)$ and it represents the wave envelope. The subscripts denote the partial differentiation of the respective variable and $\gamma$ is a real parameter. The restriction $\gamma=0$, Eq. (\ref{s21}) gives the NLS equation.
	
	The Lax pair associated with fourth-order NLS Eq. (\ref{s21}) reads
	\begin{subequations}
		\label{11}
		\begin{align}
			\Phi_x= \tilde{M}(\lambda,q) \Phi, \quad \tilde{M}(\lambda,q) = \begin{pmatrix}
				\lambda & q \\
				-\bar{q} & -\lambda
			\end{pmatrix},
			\label{s22}\\
			\Phi_t=\tilde{N}(\lambda,q)  \Phi,\quad 
			\tilde{N}(\lambda,q)= i\begin{pmatrix}
				A_1 & B_1\\
				C_1 & -A_1
			\end{pmatrix},
			\label{s23}
		\end{align} 
	\end{subequations}
	where $\Phi=(\Phi_1,\Phi_2)^T$, and
	\begin{align}
		A_1=&\frac{|q|^2}{2}+\lambda^2+\gamma(3|q|^4+\bar{q}q_{xx}+q\bar{q}_{xx}-|q_x|^2-2\lambda q \bar{q}_x\nonumber\\&+2\lambda q_x\bar{q}+4|q|^2\lambda^2+8\lambda^4),\nonumber\\
		B_1=&\frac{q_x}{2}+\lambda q+\gamma(8\lambda^3 q+4\lambda^2 q_x+6|q|^2 q_x+q_{xxx}+2\lambda q_{xx}\nonumber\\&+4\lambda|q|^2q),\nonumber\\
		C_1=&\frac{\bar{q}_x}{2}-\lambda \bar{q}+\gamma(-8\lambda^3\bar{q}+4\lambda^2\bar{q}_x+6|q|^2\bar{q}_x+\bar{q}_{xxx}-2\lambda\bar{q}_{xx}\nonumber\\&-4\lambda|q|^2\bar{q}).\nonumber
	\end{align}
	In Eq. (\ref{11}), the spectral parameter $\lambda$ is a complex one. One can unambiguously verify that the compatibility condition of the linear Eq. (\ref{11}), say $\tilde{M}_t-\tilde{N}_x+[\tilde{M},\tilde{N}]=0$, where the $[,]$ represents the commutator matrix, gives the fourth-order NLS Eq. (\ref{s21}). 
	
	
	
	\section{Instability rates of single-periodic waves for Eq. (\ref{s21})}
	To start, we evaluate the instability rate of single-periodic waves of Eq. (\ref{s21}). In this connection, we derive Jacobian elliptic function solution for Eq. (\ref{s21}). Then we determine the eigenvalues of these solutions. Subsequently, we add a linear perturbation to this solution and study the stability spectrum from the solutions of the perturbed part.
	\subsection{Single-periodic wave solutions and eigenvalues of Eq. (\ref{s21})}
	We seek periodic wave solutions of Eq. (\ref{s21}) in the following manner,
	\begin{align}
		q(x,t)=f(x)~ e^{2ibt},
		\label{s26}
	\end{align}
	where $f(x)$ is a real periodic function (which we intend to find in terms of $dn$- and $cn$- Jacobian elliptic functions) and the term $b$ in the exponential function corresponds to a real constant. Inserting Eq. (\ref{s26}) and its derivatives in Eq. (\ref{s21}), we arrive at
	\begin{align}
		f_{xx}+2|f|^2f-4bf+2\gamma(f_{xxxx}+6\bar{f}f_x^2+4|f_x|^2f\nonumber\\+8|f|^2f_{xx}+2f^2\bar{f}_{xx}
		+6|f|^4f)=0.
		\label{s27}
	\end{align}
	
	One can check that Eq. (\ref{s27}) admits $dn$- and $cn$- solutions for the following choices of $b$:
	\begin{align}
		f(x)=&~\text{dn}(x,k),\quad b=\frac{2-k^2}{4}+\frac{\gamma(6-6k^2+k^4)}{2},
		\label{s28}\\
		f(x)=&~k~\text{cn}(x,k),\quad  b=\frac{2k^2-1}{4}+\frac{\gamma(1-6k^2+6k^4)}{2},
		\label{s29}
	\end{align}
	\begin{figure*}[!ht]
		\begin{center}
			\begin{subfigure}{0.4\textwidth}
				\includegraphics[width=\linewidth]{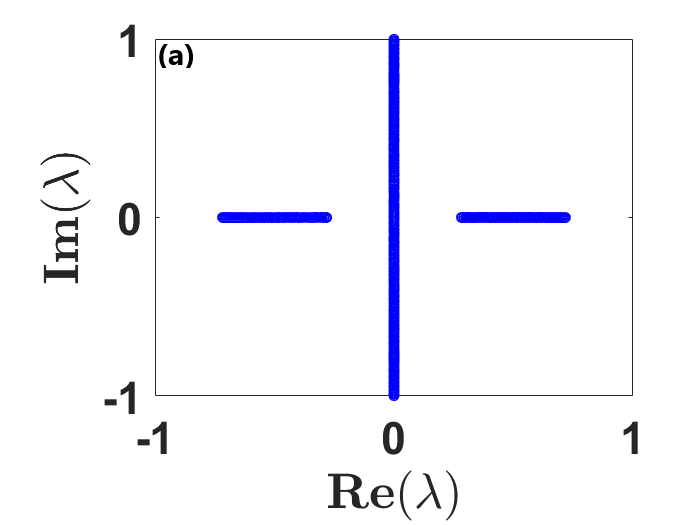}
			\end{subfigure}
		\begin{subfigure}{0.4\textwidth}
			\includegraphics[width=\linewidth]{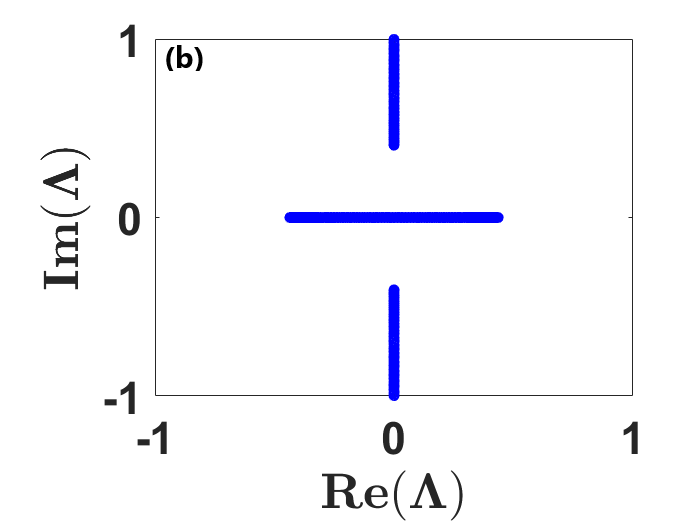}
		\end{subfigure}
		\end{center}
		\vspace{-0.3cm}
		\caption{Lax and stability spectrum for $dn$-periodic waves given in Eq. (\ref{s28}) for $k=0.9$ with $\Lambda=2\Gamma$: (a) Lax spectrum which is computed from the spatial part of the Lax pair and (b) Stability spectrum is computed through the temporal part of the Lax pair.}
		\label{fig11}
	\end{figure*}
	in which the parameter $k$ is nothing but elliptic modulus and it varies from $0$ to $1$. For $k=0$,  the  $dn$- solution gives a constant background wave $q(x,t)=e^{\frac{1}{2}(1+6\gamma)}$, whereas the $cn$- solution attains the zero background. For $k=1$, both the solutions take solitonic profile (which is normalized), $q(x,t)=\text{sech}(x)~e^{\frac{1}{2}(1+6\gamma)}$. 
	
	The functions reported in Eqs. (\ref{s28}) and (\ref{s29}) are also solutions for following two nonlinear ordinary differential equations, namely
	\begin{align}
		\frac{d^2f}{dx^2}+2|f|^2f=b_0 f,\quad \left|\frac{df}{dx}\right|^2+|f|^4=b_0 |f|^2+b_1,
		\label{s210}
	\end{align}
	where $b_0$ and $b_1$ are real constants. In the $dn$- solution (\ref{s28}), we consider $b_0=2-k^2$ and $b_1=k^2-1$, while for the $cn$- solution (\ref{s29}), we consider $b_0=2k^2-1$ and $b_1=k^2(1-k^2)$. 
	
	Since the role of eigenvalue is most important to compute the instability of both $dn$- and $cn$- periodic waves, in the following, we determine for which eigenvalues the constructed waves emerge as the solutions satisfying the Lax pair (\ref{11}). 
	
	\begin{figure*}[!ht]
		\begin{center}
			\begin{subfigure}{0.4\textwidth}
				\includegraphics[width=\linewidth]{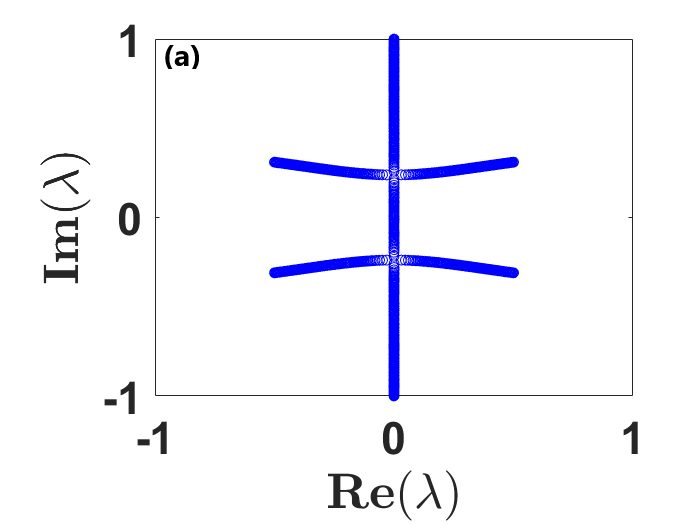}
			\end{subfigure}
			\begin{subfigure}{0.4\textwidth}
			\includegraphics[width=\linewidth]{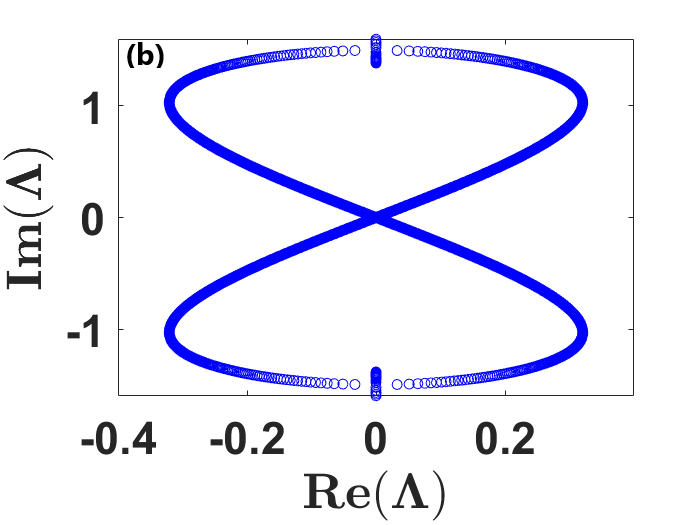}
		\end{subfigure}\\
		\begin{subfigure}{0.4\textwidth}
		\includegraphics[width=\linewidth]{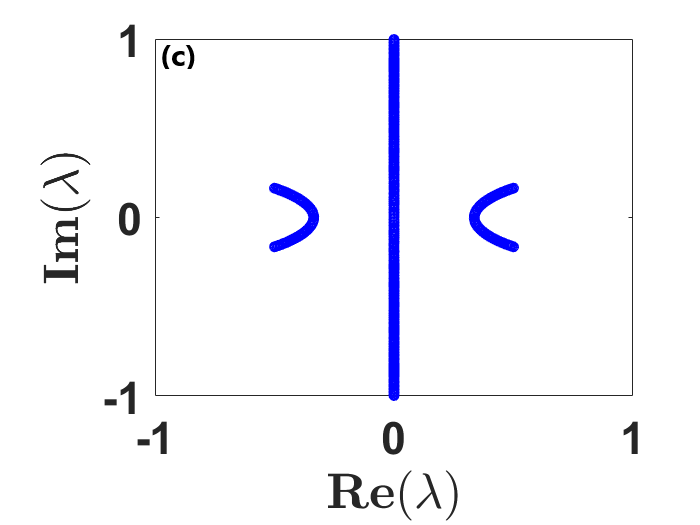}
	\end{subfigure}
\begin{subfigure}{0.4\textwidth}
	\includegraphics[width=\linewidth]{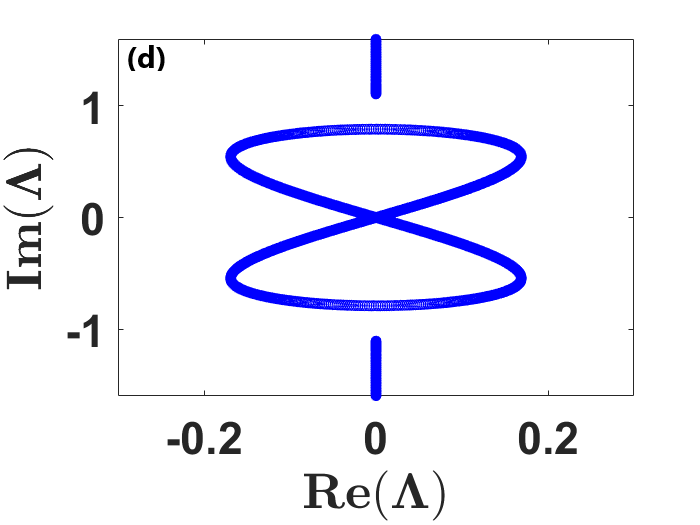}
\end{subfigure}
		\end{center}
		\vspace{-0.3cm}
		\caption{Lax (computed from the spatial part of the Lax pair) and stability (which is computed through the temporal part of the Lax pair) spectrum for $cn$-periodic waves which is given in Eq. (\ref{s29}) with $\Lambda=2\Gamma$: (a)-(b) Lax  and Stability  spectrum for $k=0.85$ and (c)-(d) Lax and Stability spectrum for $k=0.95$.}
		\label{fig22}
	\end{figure*}
	
	Now, we compute the rate of instability of single - periodic waves given in Eq. (\ref{s26}) from the spectral problem (\ref{s22}) and (\ref{s23}). 
	
	Let us generalize the solution (\ref{s26}) as
	\begin{align}
		\Phi_1(x,t)=\vartheta_1(x)~e^{(\Gamma+ib) t},\quad \Phi_2(x,t)=\vartheta_2(x)~e^{(\Gamma-ib) t},
		\label{s211}
	\end{align}
	with $\Gamma$ is a spectral parameter and $\vartheta=(\vartheta_1,\vartheta_2)^T$. Substituting Eq. (\ref{s211}) into the Lax pair (\ref{11}), the latter equation becomes
	\begin{subequations}
		\label{1}
		\begin{align}
			\vartheta_x&=\begin{pmatrix}
				\lambda & f\\
				-\bar{f} & -\lambda
			\end{pmatrix}\vartheta,
			\label{s212}\\
			\Gamma \vartheta&=i\begin{pmatrix}
				A_1-b & B_1\\ 
				C_1 & -A_1+b
			\end{pmatrix}\vartheta,
			\label{s213}
		\end{align}
	\end{subequations}
	with
	\begin{align}
		A_1=&\frac{|f|^2}{2}+\lambda^2+\gamma(3|f|^4+\bar{f}f_{xx}+f\bar{f}_{xx}-|f_x|^2-2\lambda f \bar{f}_x\nonumber\\&+2\lambda f_x\bar{f}+4|f|^2\lambda^2+8\lambda^4),\nonumber\\
		B_1=&\frac{f_x}{2}+\lambda f+\gamma(8\lambda^3 f+4\lambda^2 f_x+6|f|^2 f_x+f_{xxx}+2\lambda f_{xx}\nonumber\\&+4\lambda|f|^2f),\nonumber\\
		C_1=&\frac{\bar{f}_x}{2}-\lambda \bar{f}+\gamma(-8\lambda^3\bar{f}+4\lambda^2\bar{f}_x+6|f|^2\bar{f}_x+\bar{f}_{xxx}-2\lambda\bar{f}_{xx}\nonumber\\&-4\lambda|f|^2\bar{f})\nonumber.
	\end{align}

	We may say $\lambda$ is associated with the eigenvalue problem's Lax spectrum, Eq. (\ref{s212}), if $\vartheta\in {\omega}^\infty_1 (\mathbb{R})$. Since $f(x)$ is a periodic function, we can impose the condition $f(x+\omega_1)=f(x)$ with the fundamental period $\omega_1>0$. In this case, we can represent the bounded solutions of the linear Eq. (\ref{s212}) as \cite{chen2} (using Floquet's theorem)
	\begin{align}
		\vartheta(x)=\tilde{\vartheta}(x)~e^{i\theta x},
		\label{s214}
	\end{align}
	where $\tilde{\vartheta}(x+\omega_1)=\tilde{\vartheta}(x)$ and $\theta\in [-\frac{\pi}{\omega_1},\frac{\pi}{\omega_1}]$. The above bounded solutions (\ref{s214}) are periodic  and anti-periodic when we consider $\theta=0$ and $\theta=\pm\frac{\pi}{\omega_1}$ respectively \cite{chen2}.
	\begin{figure*}[!ht]
		\begin{center}
			\begin{subfigure}{0.4\textwidth}
				\includegraphics[width=\linewidth]{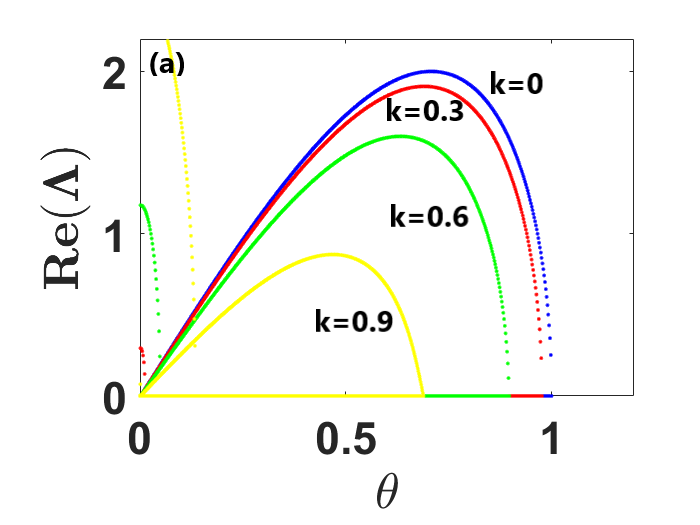}
			\end{subfigure}
		\begin{subfigure}{0.4\textwidth}
			\includegraphics[width=\linewidth]{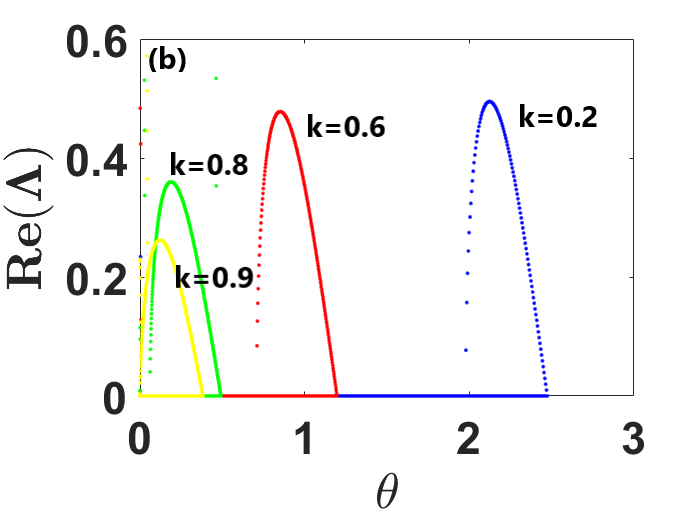}
		\end{subfigure}
		\end{center}
		\vspace{-0.3cm}
		\caption{ Instability rate for the periodic waves for different elliptic modulus values with $\Lambda=2\Gamma$: (a) $dn$-periodic waves (blue, red, green, yellow colors denote the different values of $k$, which is $0$, $0.3$, $0.6$ and $0.9$ respectively),  (b) $cn$-periodic waves (blue, red, green, yellow colors represents the four different values of $k$, which is $0.2$, $0.6$, $0.8$ and $0.9$ respectively). }
		\label{fig221}
	\end{figure*}

	From the spectral problem, Eq. (\ref{s213}), a set of linear algebraic equations can be obtained. A nonzero solution for these equations are found only when the determinant of the coefficient matrix is zero. This condition provides a connection between $\Gamma$ and $\lambda$, that is
	\begin{align}
		\Gamma^2+J(\lambda)=0,
		\label{s215}
	\end{align}
	with $J(\lambda)=4\tilde{J}(\lambda)$ and 
	\begin{align}
		\tilde{J}(\lambda)=\lambda^4-2b\lambda^2+b^2+2d.
		\label{s216}
	\end{align}
	The parameter $d$ is given by
	\begin{align}
		|f_x|^2+|f|^4-4b|f|^2=8d.
		\label{s217}
	\end{align}
	
	Now comparing Eq. (\ref{s217}) with the second equation in (\ref{s210}), we can fix $b_0=4b$ and $b_1=8d$. 
	
	The polynomial $\tilde{J}(\lambda)$ is rewritten as
	\begin{align}
		\tilde{J}(\lambda)=\lambda^4-\frac{1}{2}(f_1^2+f_2^2)\lambda^2+\frac{1}{16}(f_1^2-f_2^2),
		\label{s218}
	\end{align}
	where we have considered $4b=f_1^2+f_2^2$ and $8d=-f_1^2f_2^2$. The parameters $b$ and $d$ can be rewritten in terms of $b_0$ and $b_1$, by choosing $b_0=f_1^2+f_2^2$, $b_1=-f_1^2f_2^2$. Two positive roots of the polynomial $\tilde{J}(\lambda)$ are 
	\begin{align}
		\lambda_1=\frac{1}{2}(f_1+f_2), \quad \lambda_2=\frac{1}{2}(f_1-f_2).
		\label{s219}
	\end{align}
	Using Eq. (\ref{s219}), we can rewrite Eq. (\ref{s218}) in the form
	\begin{align}
		\tilde{J}(\lambda)=(\lambda^2-\lambda_1^2)(\lambda^2-\lambda_2^2).
		\label{s220}
	\end{align}
	
	Upon substituting Eqs. (\ref{s26}) and (\ref{s211}) in Eq. (\ref{s22}), we can obtain the value of $\lambda$. The spectral problem now takes the form,	\begin{equation}
		\begin{pmatrix} \frac{d}{dx}+i\theta & -f\\ -\bar{f} & -\frac{d}{dx}-i\theta \end{pmatrix}\begin{pmatrix} \hat{\vartheta}_1\\ \hat{\vartheta}_2 \end{pmatrix}=\lambda \begin{pmatrix} \hat{\vartheta}_1\\ \hat{\vartheta}_2 \end{pmatrix},
		\label{s221}
	\end{equation}
	where $\theta\in(0,\frac{\pi}{\omega_1})$. To compute the Lax spectrum we have to determine the eigenvalues of Eq. (\ref{s221}). For this, we solve Eq. (\ref{s221}) numerically using Floquet-Bloch decomposition method \cite{chen1}.
	\subsection{Linear perturbation to the single-periodic wave (\ref{s26})}
	Upon perturbing the solution  (\ref{s26}) linearly,
	\begin{align}
		q(x,t)=[f(x)+g(x,t)]~e^{2ibt},
		\label{s223}
	\end{align}
	and inserting the latter expression into Eq. (\ref{s21}) and leaving out the quadratic terms in $g$, we arrive at
	\begin{subequations}
		\label{s224}
		\begin{align}
			\label{i11}
			&ig_t-2bg+g_{xx}+4|f|^2g+2f^2\bar{g}+\gamma( 18|f|^4g+12|f|^2f^2\bar{g}\nonumber\\&+6f_x^2\bar{g}+4|f_x|^2g	+12\bar{f}f_xg_x+4f \bar{f}_xg_x+8\bar{f}f_{xx}g+8f f_{xx}\bar{g}\nonumber\\&+4f\bar{f}_{xx}g+4f f_x\bar{g}_x+8|f|^2g_{xx}+2f^2\bar{g}_{xx}+g_{xxxx})=0,\\\nonumber\\
			&-i\bar{g}_t-2b\bar{g}+\bar{g}_{xx}+4|f|^2\bar{g}+2\bar{f}^2g+\gamma(18|f|^4\bar{g}+12|f|^2\bar{f}^2g\nonumber\\&+6\bar{f}_x^2g+4|f_x|^2\bar{g}+12f\bar{f}_x\bar{g}_x+4\bar{f} f_x\bar{g}_x+8f\bar{f}_{xx}\bar{g}+8\bar{f} \bar{f}_{xx}g\nonumber\\&+4\bar{f}f_{xx}\bar{g}+4\bar{f} \bar{f}_xg_x+8|f|^2\bar{g}_{xx}+2\bar{f}^2g_{xx}+\bar{g}_{xxxx})=0.
		\end{align}
	\end{subequations}
	
	We assume a simple nontrivial separable solution of Eq. (\ref{s224}) in the form
	\begin{align}
		g(x,t)=m_1(x)~e^{ \Lambda t}, \quad \bar{g}(x,t)=m_2(x)~e^{\Lambda t},
		\label{s225}
	\end{align}
	where $m_1$ and $m_2$ are functions of $x$, $\Lambda$ represents the spectral parameter and $m=(m_1,m_2)^{T}$. Using Eqs. (\ref{s224}) and (\ref{s225}) we rewrite the spectral stability problem as,
	\begin{align}
		\label{s226}
		i\Lambda\sigma_3 ~m+\begin{pmatrix} s_{11} & s_{21}\\ \bar{s}_{21} & \bar{s}_{11} \end{pmatrix}m=0, \quad
		\sigma_3 = \begin{pmatrix}
			1 & 0 \\
			0 & -1
		\end{pmatrix},
	\end{align}
	with
	\begin{subequations}
		\label{s227}
		\begin{align}
			s_{11}=&-2b+4|f|^2+\partial_{xx}+\gamma(18|f|^4+4|f_x|^2+12\bar{f}f_x\partial_x\nonumber\\&+4f\bar{f}_x\partial_x+8f \bar{f}_{xx}+4f \bar{f}_{xx}+8|f|^2\partial_{xx}+\partial_{xxxx}), \\ 
			s_{21}=&2f^2+\gamma(12|f|^2f^2+6f_x^2+4ff_x\partial_x+8f f_{xx}+2f^2\partial_{xx}).
		\end{align}
	\end{subequations}

	In Eqs. (\ref{s226}) with (\ref{s227}), $\Lambda$ belongs to the stability spectrum, given that $m$ is real and the eigenvalue problem (\ref{1}) for the  Lax spectrum is related to the eigenvalue $\lambda$.  $\vartheta^2_1$ and $\vartheta^2_2$ are the bounded squared eigenfunctions that determines the eigenfunctions of Eq. (\ref{s226}), which are also bounded, that is $m_1=\vartheta^2_1$ and $m_2=-\vartheta^2_2$. Moreover, we also come across a relation $\Lambda=2\Gamma$ during this process, with $\Gamma$ calculates the eigenvalues. For the calculated eigenvalue, $\lambda$ in the Lax spectrum, if $Re(\Lambda)>0$ then the single-periodic waves given in Eq. (\ref{s26}) is said to be spectrally unstable. In the eigenvalue spectrum (on the $\Lambda$-plane), if $Re(\Lambda)>0$ passing through the origin $(0,0)$ and is horizontal to the imaginary axis, then the periodic waves are said to be modulationally unstable. It has been pointed out in the literature that, RWs will degenerate into propagating algebraic solitons instead of localization, when the periodic waves are unstable spectrally and stable modulationally (see Ref. \cite{pelinovsky1}). On the other hand, if the spectrum with $Re(\Lambda)>0$ in the $\Lambda$-plane intersects the origin and it vertically reaches the imaginary axis, then the corresponding RWs reduces into an algebraic soliton which propagates \cite{chen2}.
	
	Now, we analyze the modulational and spectral stability of the single-periodic waves. Substituting the relation $\Lambda=2\Gamma$ in the expression $\Gamma^2+4\tilde{J}(\lambda)=0$, we obtain $\Lambda^2 + 16 \tilde{J}(\lambda)=0$. The eigenspectrum of the perturbed part and the unperturbed part are connected by this relation. This enables us to calculate the growth rate of instability with the help of the relation $\Lambda=\pm 4i\sqrt{\tilde{J}(\lambda)}$. Hence, we can inspect the change in the growth rate of instability of both $dn$- and $cn$- periodic waves by varying $k$.
	
	The Lax and stability spectrum of Eq. (\ref{1}) is shown in Fig. \ref{fig11} for $k=0.9$ by considering the $dn$-periodic wave with $f_1=1$  and $f_2=\sqrt{1-k^2}$. Using Eq. (\ref{s221}) we compute the Lax spectrum, which depends on the $\lambda$ plane whose outcome is demonstrated in Fig. \ref{fig11}(a). The unstable spectrum is calculated from the expression, $\Lambda=\pm 4i\sqrt{\tilde{J}(\lambda)}$ and is displayed in Fig. \ref{fig11}(b), where $\tilde{J}(\lambda)$ is given in Eq. (\ref{s220}). In Fig. \ref{fig11}(b), the finite segment line that occur in real part $(Re(\Lambda))$ touches the origin. From this, we conclude that the $dn$-periodic wave solution (\ref{s28}) is both spectrally and modulationally unstable.
	
	For the $cn$-periodic wave, we take $f_1=1$ and $f_2=i\sqrt{1-k^2}$.
	The Lax spectrum of Eq. (\ref{s212}) for the $cn$-periodic wave is given in Figs. \ref{fig22}(a) and \ref{fig22}(c) for $k=0.85$ and $k=0.95$, respectively. Figures \ref{fig22}(b) and \ref{fig22}(d) represent the stability spectrum (unstable) of $cn$-periodic waves. Despite the changes in the Lax spectrum, the stability spectrum of both the cases (for two different $k$ values) turns out to be similar. The only difference that we notice in these figures is that the eight-bands, which are connected both in real and imaginary axis and the purely imaginary bands (real part is zero) intersect for $k=0.85$ (Fig. \ref{fig22}(b)), whereas they do not intersect for $k=0.95$ (Fig. \ref{fig22}(d)). From the outcome, we conclude that the $cn$-periodic wave given in Eq. (\ref{s29}) is spectrally and modulationally unstable.

	Figure \ref{fig221} shows the rate of the instability of single-periodic waves for different elliptic modulus ($k$) values. In Fig. \ref{fig221}, we plot $Re(\Lambda)$ against $\theta$ where $\theta$ is the Floquet parameter which we vary from $0$ to $\frac{\pi}{\omega_1}$. The largest maximal instability is obtained for $k=0$ as shown in Fig. \ref{fig221}(a). When increasing the value of $k$ from $0$ to $1$ the largest maximal instability decreases. For $k=1$ it vanishes. Figure \ref{fig221}(b) represents the rate of instability of the $cn$-periodic waves, which is given in Eq. (\ref{s29}). The maximal instability decreases when we vary the value of $k$ from $0$ to $0.9$. In other words, the highest maximal instability  is observed for $k=0.2$. Thus, the largest maximal instability of $dn$-periodic wave is larger than that of $cn$-periodic wave. The maximal instability rate of single-periodic wave is larger for the fourth-order NLS equation when compared to the standard NLS equation (see Ref. \cite{pelinovsky}).
	
	\section{Instability rates of double-periodic waves for Eq. (\ref{s21})}
	
	In the following, we evaluate the rate of instability of waves that are periodic both spatially and temporally and compare the rate of instability of these double-periodic waves with single-periodic waves. Equation (\ref{s21}) admits the following form of double-periodic solution
	\begin{align}
		q(x,t)=[f(x,t)+i \Delta(t)]e^{i\theta(t)}.
		\label{s41}
	\end{align}
	In Eq. (\ref{s41}), the first term $f(x,t)$ is periodic in both $x$ and $t$, whereas the second term $\Delta(t)$ should be a double-period one in time $t$. The term $\theta(t)$ in the exponential is a real function, which is constant in the variable $x$. The solutions of this kind of nature for the Eq. (\ref{s21}) have already been reported \cite{crabb} and they read as	
		\begin{eqnarray}
			q(x,t)=k\frac{\text{cn}(Bt,k)~\text{cn}(\sqrt{\nu}x,\tau)+i\sqrt{\nu}~\text{sn}(Bt,k)~\text{dn}(\sqrt{\nu}x,\tau)}{\sqrt{\nu}~\text{dn}(\sqrt{\nu}x,\tau)-\text{dn}(B t,k)~\text{cn}(\sqrt{\nu}x,\tau)}e^{i\zeta t},\quad \tau=\frac{\sqrt{1-k}}{\sqrt{\nu}},
			\label{s42}\\
			q(x,t)=\frac{\text{dn}(Bt,k)~\text{cn}(\sqrt{2}x,\tau)+i\sqrt{\nu k}~\text{sn}(Bt,k)}{\sqrt{\nu}-\sqrt{k}~\text{cn}(Bt,k)~\text{cn}(\sqrt{2}x,\tau)}e^{i\zeta k t}, \quad \quad \quad \quad \quad~ \tau=\frac{\sqrt{1-k}}{\sqrt{2}}.
			\label{s43}
		\end{eqnarray}
	where $\nu=1+k$ and the constants $B$ and $\zeta$ that appear in the Eqs. (\ref{s42}) and (\ref{s43}) are related to the system parameter $\gamma$ through the relations, $B=1+8\gamma$ and $\zeta=1+(8-\frac{2}{k^2})\gamma$ \cite{crabb}. 
	
	Equations (\ref{s42}) and (\ref{s43}) also satisfy the following identities
	\begin{align}
		q(x,t)=&~\Phi(x,t)~e^{2i bt},\nonumber\\ \Phi(x+\omega_1,t)=&~\Phi(x,t+\omega_2)=\Phi(x,t),
		\label{s44}
	\end{align}
	where $\omega_1>0$ and $\omega_2>0$ are respectively the fundamental period in space and time coordinate and one should consider $2b=\zeta$ for Eq. (\ref{s42}) and $2b=\zeta k$ for Eq. (\ref{s43}).
	\begin{figure*}[!ht]
		\begin{center}
			\begin{subfigure}{0.4\textwidth}
				\includegraphics[width=\linewidth]{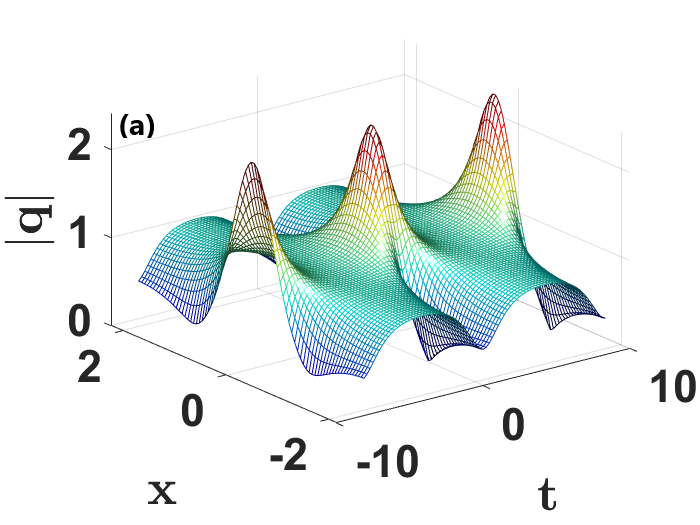}
			\end{subfigure}
			\begin{subfigure}{0.4\textwidth}
			\includegraphics[width=\linewidth]{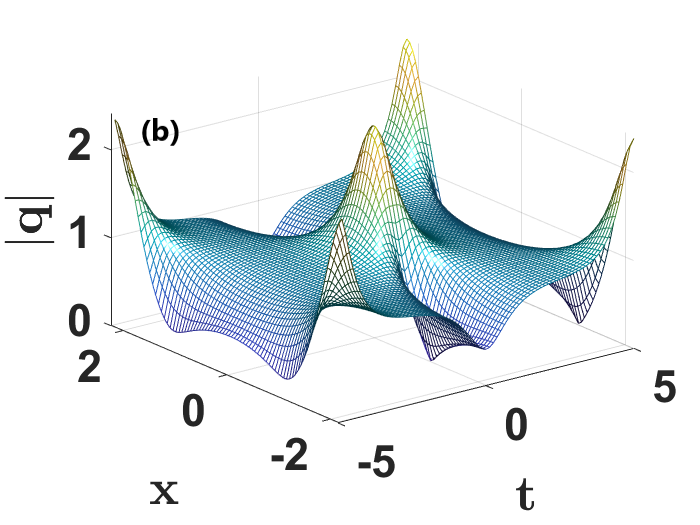}
		\end{subfigure}
		\end{center}
		\vspace{-0.3cm}
		\caption{ Double-periodic waves for $k=0.999$ with the system parameter $\gamma=0.01$: (a) Eq. (\ref{s42}) gives the phase-repeated wave patterns and (b) Eq. (\ref{s43}) gives phase-alternating patterns.} 
		\label{fig1}
	\end{figure*}
	
	
	The double-periodic solutions given in (\ref{s42}) and (\ref{s43}) can be generalized to the form given in (\ref{s44}). Here, $\Phi$ represents the solution to the linear equation (\ref{11}) along with
	\begin{align}
		\Phi_1(x,t)=&~\vartheta_1(x,t)~e^{\eta x+(\Gamma+i b ) t},\nonumber\\ \Phi_2(x,t)=&~\vartheta_2(x,t)~e^{\eta x+(\Gamma-i b ) t},
		\label{s45}
	\end{align}
	where $\eta$, $\Gamma$ are spectral parameters and $\vartheta=(\vartheta_1,\vartheta_2)^T$. Substituting Eq. (\ref{s45}) into the Lax pair Eq. (\ref{11}) we obtain the following set of equations, namely
	\begin{subequations}
		\label{s467}
		\begin{align}
			\vartheta_x+\eta \vartheta=\begin{pmatrix}
				\lambda & \Phi\\
				-\bar{\Phi} & -\lambda
			\end{pmatrix}\vartheta,
			\label{s46}\\
			\vartheta_t+\Gamma\vartheta=i\begin{pmatrix}
				\hat{A}_1- b & \hat{B}_1\\ 
				\hat{C}_1 & -\hat{A}_1+ b
			\end{pmatrix}\vartheta,
			\label{s47}
		\end{align}
	\end{subequations}
	where,
	\begin{align}
		\hat{A}_1=&\frac{|\Phi|^2}{2}+\lambda^2+\gamma(3|\Phi|^4+\bar{\Phi}\Phi_{xx}+\Phi\bar{\Phi}_{xx}-|\Phi_x|^2\nonumber\\&-2\lambda \Phi \bar{\Phi}_x+2\lambda \Phi_x\bar{\Phi}+4|\Phi|^2\lambda^2+8\lambda^4),\nonumber\\
		\hat{B}_1=&\frac{\Phi_x}{2}+\lambda \Phi+\gamma(8\lambda^3 \Phi+4\lambda^2 \Phi_x+6|\Phi|^2 \Phi_x+\Phi_{xxx}\nonumber\\&+2\lambda \Phi_{xx}+4\lambda|\Phi|^2\Phi),\nonumber\\
		\hat{C}_1=&\frac{\bar{\Phi}_x}{2}-\lambda \bar{\Phi}+\gamma(-8\lambda^3\bar{\Phi}+4\lambda^2\bar{\Phi}_x+6|\Phi|^2\bar{\Phi}_x+\bar{\Phi}_{xxx}\nonumber\\&-2\lambda\bar{\Phi}_{xx}-4\lambda|\Phi|^2\bar{\Phi}).\nonumber
	\end{align}
	
	The parameters $\eta$ and $\Gamma$ are independent of space ($x$) and time ($t$) coordinates.
	Upon considering the function $q(x,t)$ in Eq. (\ref{s44}), the compatibility condition of the system of linear equations ((\ref{s22}) and (\ref{s23})) satisfies the fourth-order NLS Eq. (\ref{s21}).
	The spectral parameters $\eta$ and $\Gamma$ can be fixed in terms of $\lambda$ using the periodicity condition, $\vartheta(x+\omega_1,t)=\vartheta(x,t+\omega_2)=\vartheta(x,t)$ (Floquet theorem). The condition which defines the Lax spectrum is that the eigenvalue $\lambda$ corresponds to an admissible set whereby the solution given in Eq. (\ref{s45}) is bounded in $x$ with $\eta=i\theta$, $\theta\in[-\frac{\pi}{\omega_1},\frac{\pi}{\omega_1}]$. The associated eigenvalue $\lambda$ is calculated using eigenvalue solver technique with the help of the spectral problem $(\ref{s46})$ for every $t$ which is real ($\vartheta(x+\omega_1,t)=\vartheta(x,t)$). The stability spectrum defined by the condition $\vartheta(x,t+\omega_2)=\vartheta(x,t)$ can be obtained by solving the spectral problem (\ref{s47}) for $\Gamma$ with the help of $\lambda$ in the Lax spectrum when all $x$ is real. In other words, the construction of Lax spectrum and stability spectrum of double-periodic waves depends on $x$ and $t$. The time coordinate $t$ is fixed when $x$ having a bounded value, and the space coordinate $x$ is fixed when $t$ has a bounded value from which we can compute the Lax and stability spectrum. The separation between space $(x)$ and time $(t)$ coordinates arises since we examine the stability of double-periodic waves. Otherwise, a similar procedure can be followed to determine the perturbed eigenvalue $\Lambda=2\Gamma$ as in the study of instability of the single-periodic wave with $q=\Phi$. The spectral parameter $\Gamma$ is defined by the range $Im(\Gamma)$ in $[-\frac{\pi}{\omega_1}, \frac{\pi}{\omega_1}]$. Here $Re(\Gamma)$ determines the rate of instability ($Re(\Lambda)$) through the relation $\Lambda=2\Gamma$.
	\begin{figure*}[!ht]
		\begin{center}
				\begin{subfigure}{0.4\textwidth}
				\includegraphics[width=\linewidth]{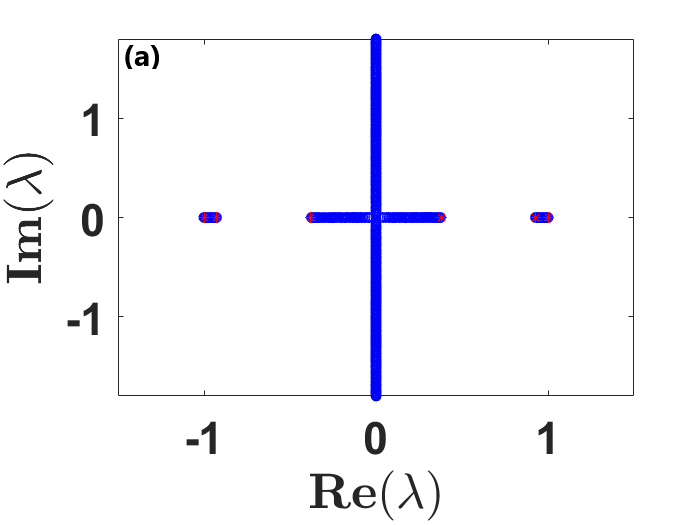}
			\end{subfigure}
			\begin{subfigure}{0.4\textwidth}
			\includegraphics[width=\linewidth]{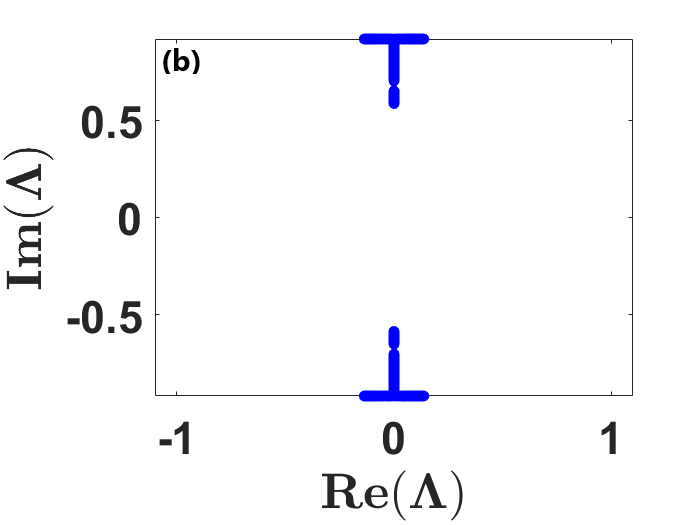}
		\end{subfigure}\\
\begin{subfigure}{0.4\textwidth}
	\includegraphics[width=\linewidth]{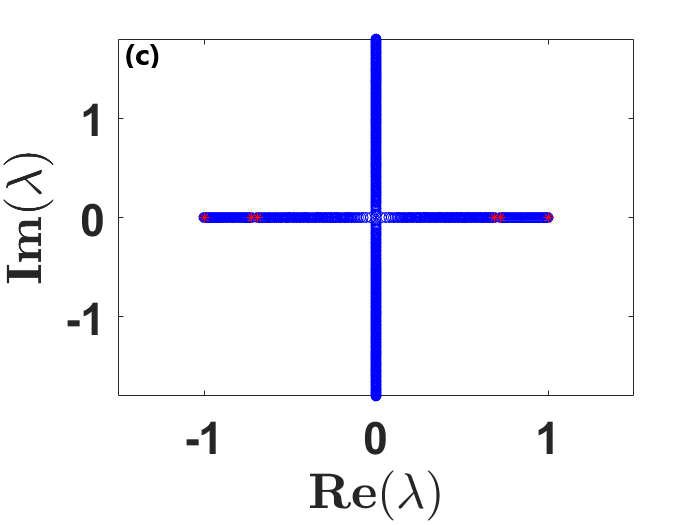}
\end{subfigure}
\begin{subfigure}{0.4\textwidth}
	\includegraphics[width=\linewidth]{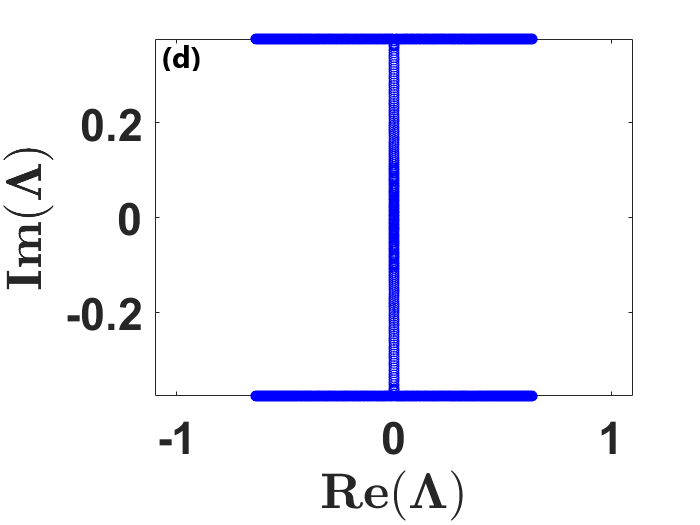}
\end{subfigure}
		\end{center}
		\vspace{-0.3cm}
		\caption{Lax (computed from the spatial part of the Lax pair) and stability (computed through the temporal part of the Lax pair) spectrum for double-periodic waves in Eq. (\ref{s42}) with $\gamma=0.01$: (a)-(b) Lax and Stability spectrum for $k=0.2$ and (c)-(d) Lax and Stability spectrum for $k=0.999$. }
		\label{fig2}
	\end{figure*}
	\begin{figure*}[!ht]
		\begin{center}
			\begin{subfigure}{0.4\textwidth}
				\includegraphics[width=\linewidth]{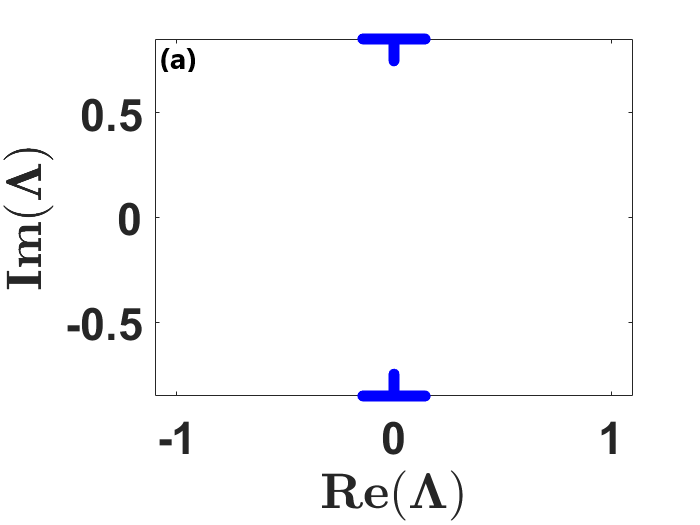}
			\end{subfigure}
		\begin{subfigure}{0.4\textwidth}
			\includegraphics[width=\linewidth]{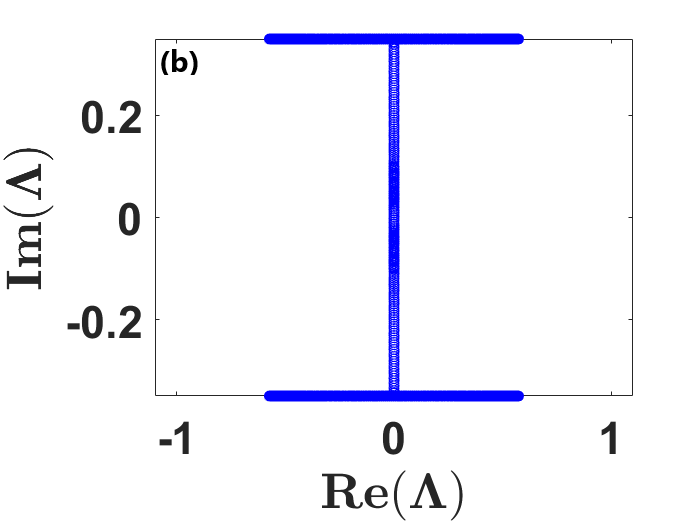}
		\end{subfigure}
		\end{center}
		\vspace{-0.3cm}
		\caption{Stability (computed through the temporal part of the Lax pair) spectrum for double-periodic waves in Eq. (\ref{s42}) when $\gamma=0$ it reduce to standard NLS solution: (a) $k=0.2$ and (b) $k=0.999$.}
		\label{fig21}
	\end{figure*}
	
	\begin{figure*}[!ht]
		\begin{center}
			\begin{subfigure}{0.4\textwidth}
				\includegraphics[width=\linewidth]{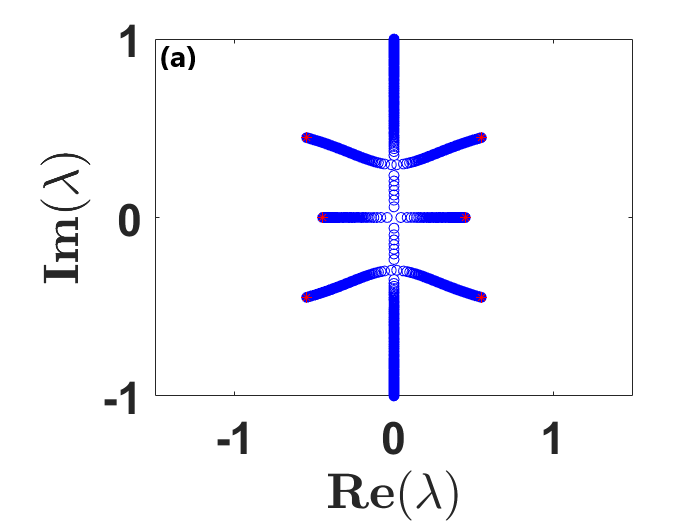}
			\end{subfigure}
			\begin{subfigure}{0.4\textwidth}
			\includegraphics[width=\linewidth]{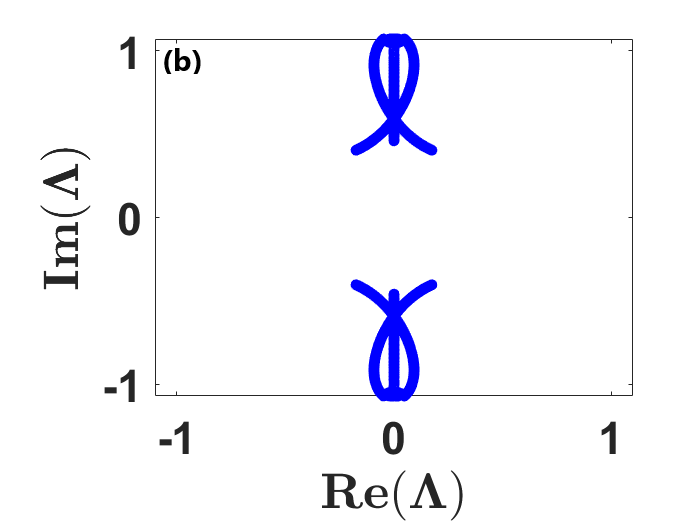}
		\end{subfigure}\\
	\begin{subfigure}{0.4\textwidth}
		\includegraphics[width=\linewidth]{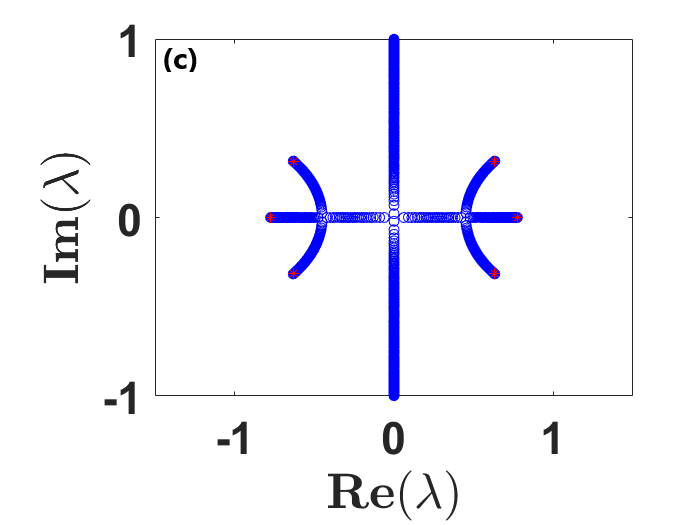}
	\end{subfigure}
	\begin{subfigure}{0.4\textwidth}
	\includegraphics[width=\linewidth]{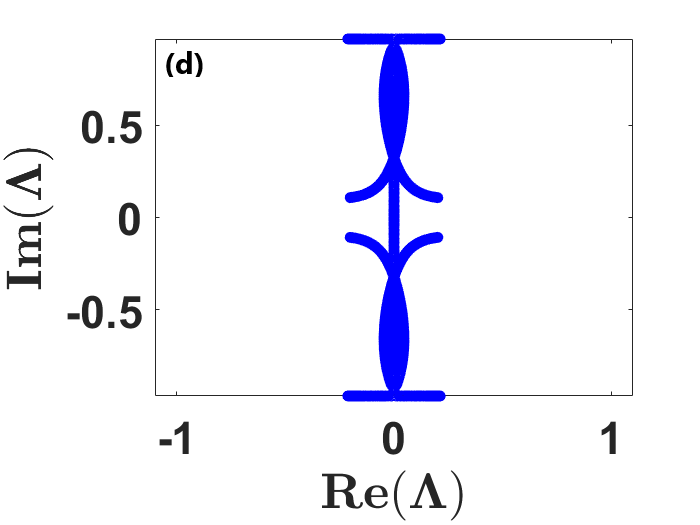}
\end{subfigure}\\
	\begin{subfigure}{0.4\textwidth}
	\includegraphics[width=\linewidth]{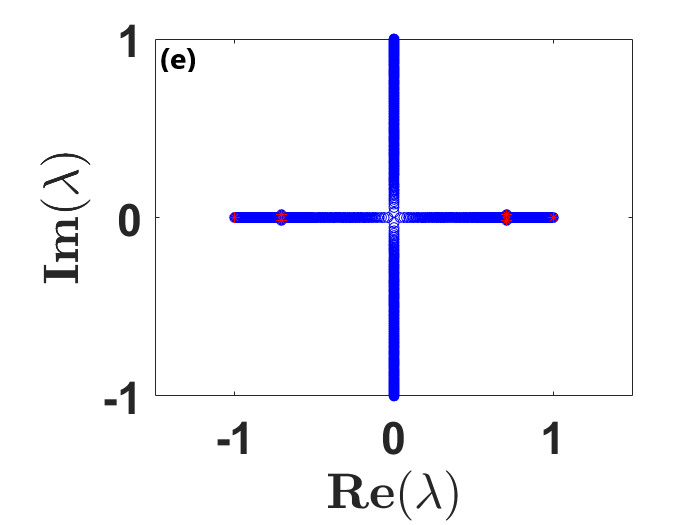}
\end{subfigure}
	\begin{subfigure}{0.4\textwidth}
	\includegraphics[width=\linewidth]{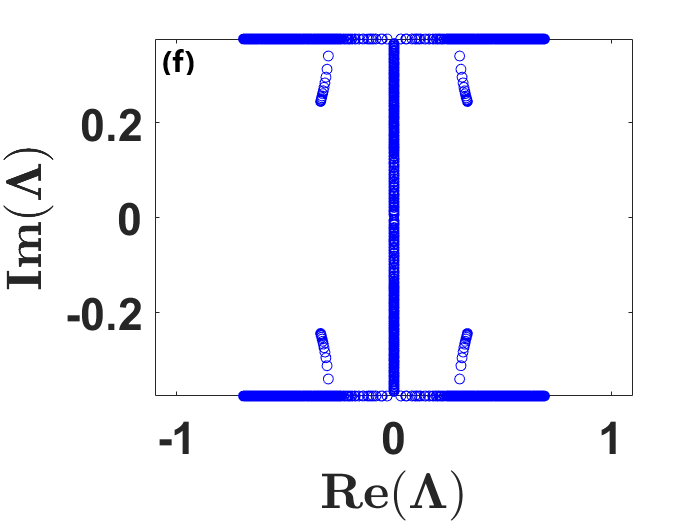}
\end{subfigure}	
		\end{center}
		\vspace{-0.3cm}
		\caption{Lax (computed from the spatial part of the Lax pair) and stability (computed through the temporal part of the Lax pair) spectrum for double-periodic waves in Eq. (\ref{s43}) with $\gamma=0.01$: (a)-(b) Lax and Stability spectrum for $k=0.2$,  (c)-(d) Lax and Stability spectrum for $k=0.6$ and (e)-(f) Lax and Stability spectrum for $k=0.999$.}
		\label{fig3}
	\end{figure*}
	\begin{figure*}[!ht]
		\begin{center}
	\begin{subfigure}{0.25\textwidth}
	\includegraphics[width=\linewidth]{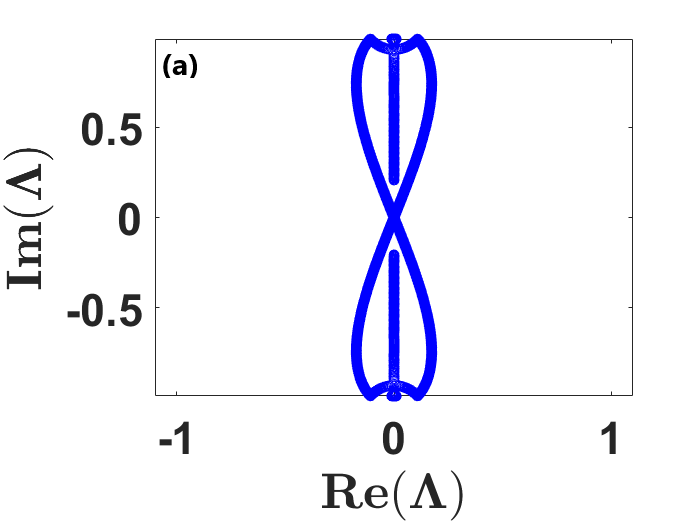}
\end{subfigure}	
	\begin{subfigure}{0.25\textwidth}
	\includegraphics[width=\linewidth]{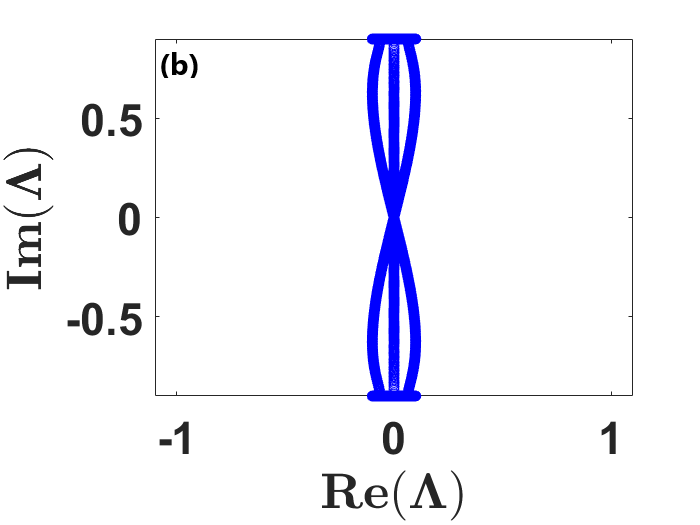}
\end{subfigure}
	\begin{subfigure}{0.25\textwidth}
	\includegraphics[width=\linewidth]{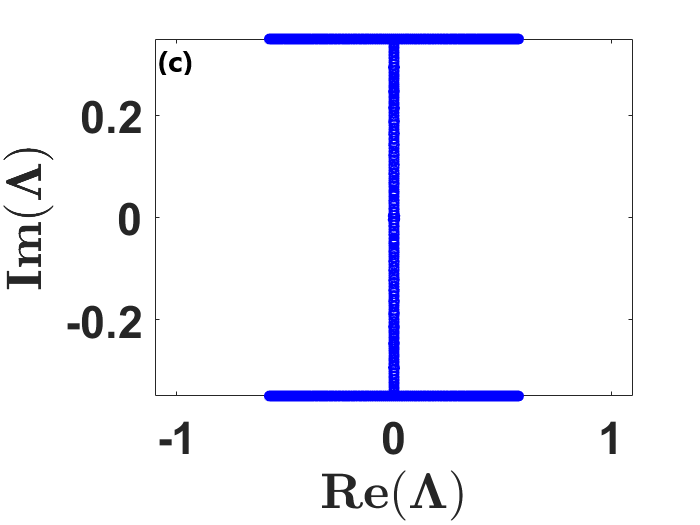}
\end{subfigure}	
		\end{center}
		\vspace{-0.3cm}
		\caption{Stability (computed through the temporal part of the Lax pair) spectrum for double-periodic waves in Eq. (\ref{s43}) when $\gamma=0$ it reduce to standard NLS solution: (a) $k=0.2$, (b) $k=0.6$ and (c) $k=0.999$. }
		\label{fig31}
	\end{figure*}
	
	\begin{figure*}[!ht]
		\begin{center}
			\begin{subfigure}{0.4\textwidth}
				\includegraphics[width=\linewidth]{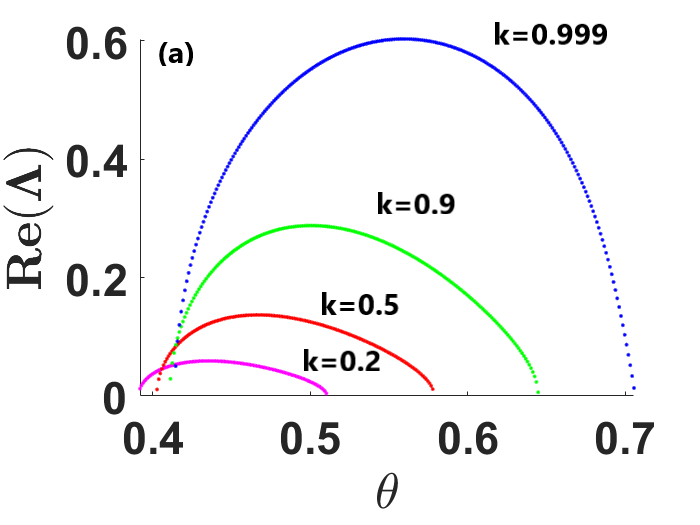}
			\end{subfigure}
			\begin{subfigure}{0.4\textwidth}
			\includegraphics[width=\linewidth]{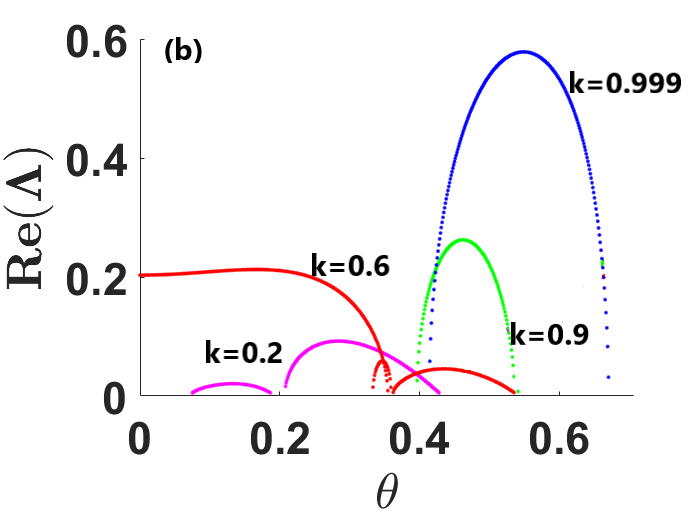}
		\end{subfigure}\\
	\begin{subfigure}{0.4\textwidth}
		\includegraphics[width=\linewidth]{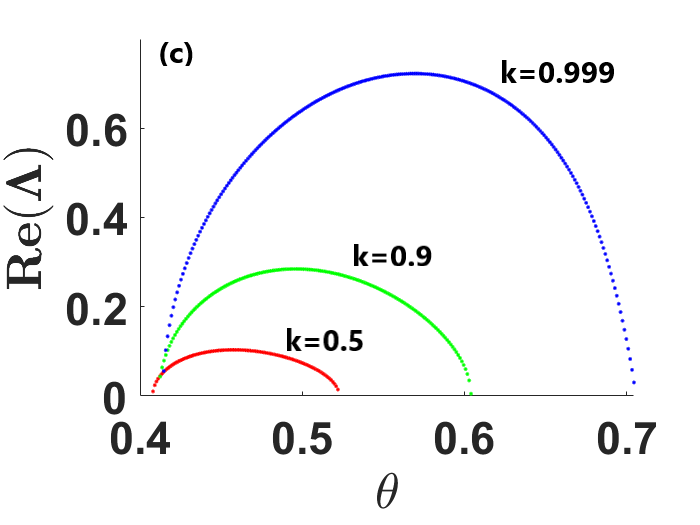}
	\end{subfigure}
	\begin{subfigure}{0.4\textwidth}
	\includegraphics[width=\linewidth]{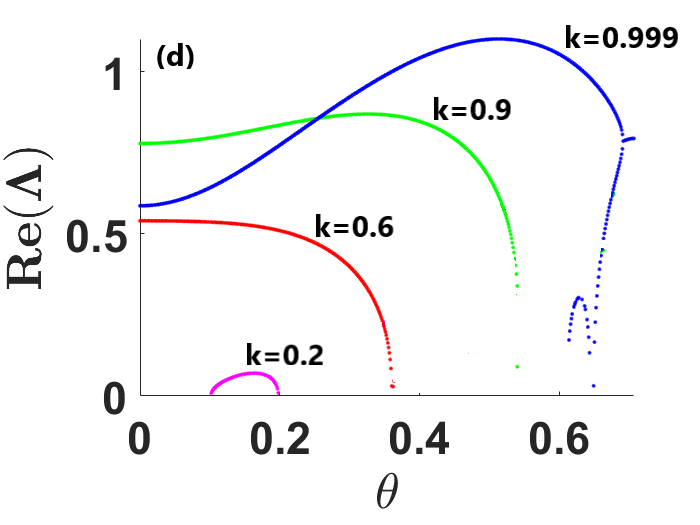}
\end{subfigure}
		\end{center}
		\vspace{-0.3cm}
		\caption{Rate of instability for the double-periodic waves for different elliptic modulus values: (a), (c) Instability rate for Eq. (\ref{s42})  with two different values of $\gamma$, which are $0.01$ and $0.05$ (blue, green, red, pink colors denote the different values of $k$, which is $0.999$, $0.9$, $0.5$ and $0.2$ respectively) and (b), (d) Instability rate for Eq. (\ref{s43})  with $\gamma=0.01$ and  $\gamma=0.05$ (blue, green, red, yellow colors denote the different values of $k$, which are $0.999$, $0.9$, $0.6$ and $0.2$ respectively).}
		\label{fig4}
	\end{figure*}
	
	In Fig. \ref{fig1}, we plot the surface plots of $q(x,t)$ of the double-periodic solutions. Figure \ref{fig1}(a) represents the amplitude of the double-periodic waves $(\ref{s42})$ for $k=0.999$. In Fig. \ref{fig1}(b), we plot the amplitude of double-periodic waves (\ref{s43}) for the same elliptic modulus ($k$) value. From these figures, one may notice that the solutions given in Eqs. (\ref{s42}) and (\ref{s43}) represent phase-repeated and phase-alternated wave patterns respectively. 
	
	If the spectrum $Re(\Lambda)>0$ in the Lax spectrum with eigenvalue $\lambda$, then Eq. (\ref{s44}), which represents the double-periodic wave solution, is said to be spectrally unstable. The Lax spectrum for the eigenvalue problem (\ref{s46}) and stability spectrum of the double-periodic solutions for two different $k$ values is displayed in Fig. \ref{fig2}. Figures \ref{fig2}(a) and \ref{fig2}(c) represent the Lax spectrum on the $\lambda$-plane for $k=0.2$ and $k=0.999$ respectively. The stability spectrum of the solution (Eq. (\ref{s42})) is demonstrated in Figs. \ref{fig2}(b) and \ref{fig2}(d) for the same values of $k$. Here we notice that for every $k$ value ($0$ to $1$), the unstable spectrum can be found at the strip's boundary $Im(\Lambda)$. Thus, the double-periodic wave given in Eq. (\ref{s42}) is spectrally unstable. The solution (\ref{s42}) of the fourth-order NLS equation converges to the double-periodic solution of the standard NLS equation when $\gamma=0$. The corresponding stability spectrum of NLS equation is demonstrated in Fig. \ref{fig21}. Figures \ref{fig21}(a) and \ref{fig21}(b) show the stability of such double-periodic solution for two different $k$ values ($0.2$ and $0.999$). These two figures help us to compare the stability spectrum of Eq. (\ref{s21}) with NLS equation. We can see that the following stability differences in the Figs. \ref{fig21} (NLS equation) and \ref{fig2} (fourth-order NLS equation). The presence of eigenvalues (spectral line) on the $Re(\Lambda)$ in the stability spectrum of Fig. \ref{fig21} is smaller compared to Fig. \ref{fig2}. Also, the spectral line on the $Im(\Lambda)$ is increased in Fig. \ref{fig2}(b) compared to NLS Fig. \ref{fig21}(a). This spectral line changes because of the higher-order dispersion parameter $\gamma$.
	
	Figure \ref{fig3} is same as Fig. \ref{fig2}. Figures \ref{fig3}(a) and \ref{fig3}(b) represent the Lax and stability spectrum  of (\ref{s43}) for $k=0.3$, whereas Figs. \ref{fig3}(c) and \ref{fig3}(d) are drawn for $k=0.6$ and Figs. \ref{fig3}(e) and \ref{fig3}(f) are depicted for $k=0.999$. The Lax spectrum has three types of bands (upper, center, lower) on the $\lambda$-plane, of which two bands (upper and lower) connect both the imaginary and real axis (Fig. \ref{fig3}(a), Fig. \ref{fig3}(c) and Fig. \ref{fig3}(e)). Whereas the third (center) band occupies the real axis alone. The spectrum which is unstable on the $\Lambda$-plane includes two symmetrically inverted heart shaped bands having a boundary at $Im(\Lambda)=\pm \frac{2\pi}{\omega_2}$. The two bands are connected with the pure imaginary band. When $k\rightarrow1$, these two bands narrow down and the stability spectrum appears similar to Fig. \ref{fig2}. This is due to the fact that, the double-periodic solutions (\ref{s42}) and (\ref{s43}) converge to one kind of Akhmediev breather (AB) solution for $k=1$. For the choice $\gamma=0$, the double-periodic solution (\ref{s43}) of Eq. (\ref{s21}) becomes the solution of NLS equation. The corresponding stability spectrum is given in Fig. \ref{fig31}. Figures \ref{fig31}(a), \ref{fig31}(b) and \ref{fig31}(c) expose the stability spectrum for corresponding to the values of $k=0.3,~0.6$ and $0.999$. Here also, we can see that the following stability differences in the Figs. \ref{fig31} (NLS equation) and \ref{fig3} (fourth-order NLS equation). We would like to point out that in the NLS equation, the two symmetrically inverted heart shaped bands (eight-band) intersects \cite{pelinovsky} at the origin $(0,0)$, whereas in the fourth-order NLS equation the two symmetrically inverted heart shaped bands (like eight-band) does not intersect at the origin (it goes away from the origin). Also, the spectal line on the $Re(\Lambda)$ in the stability spectrum of Fig. \ref{fig31} is smaller when compare to Fig. \ref{fig3}. This is because of the higher-order dispersion parameter $\gamma$.
	
	Instability rate of the fourth-order NLS equation for the aforementioned two double-periodic waves  is illustrated in Fig. \ref{fig4} for four different values of $k$. Here $Re(\Lambda)$ and the Floquet parameter $\theta$ ($\theta\in [0,\frac{\pi}{\omega_1}]$) are plotted against each other with $\eta=i\theta$. The instability rate is highest when $k=0.999$, that is, at the Akhmediev breather (when $k=1$). The double-periodic wave given in Eq. (\ref{s42}) can be visualized in Fig. \ref{fig4}(a). The unstable spectrum begins at the same value as the cut off $\theta$ and extends to $\theta=\frac{\pi}{\omega_1}$. When we decrease the value of $k$ from a higher value to a lower value, the maximal instability decreases. For $k=0$ the double-periodic solutions converges to the NLS soliton. In Fig. \ref{fig4}(b), we display the rate of instability for the double-periodic wave of (\ref{s43}), where we can see the rate of instability is high for the choice $k\rightarrow0$, at the time the double-periodic wave converges to the $cn$-periodic waves. The highest instability rate is obtained for $k=0.999$ ($k\approx1$), that is, when it reduces to the Akhmediev breather solution. Here, we can observe that, for $k=1$, the instability rate reaches its maximum level. The instability rate decreases when we decrease the value of $k$ from higher to lower ($0.999$ to $0.2$). For both the double-periodic waves, the instability rate attain its maximal level to $1$ (because of its unity). To understand the instability rate difference between NLS and (\ref{s21}), we slightly increase the system parameter $\gamma$ value from $0.01$ to $0.05$. The outcome of Eqs. (\ref{s42}) and (\ref{s43}) are illustrated in Figs. \ref{fig4}(c) and \ref{fig4}(d) with $k=0.999$. In Fig. \ref{fig4}(c), for $k=0.2$, the instability rate vanishes while we increase the value of $\gamma$ but not in the case of Fig. \ref{fig4}(d). It also checks out that the instability rate increases when the value of system parameter $\gamma$ is increased. In both the solutions, the largest instability rate gradually increases, and the width of the instability band changes when we increase the system parameter $\gamma$ value.
	\section{\label{sec5}Conclusion}
	In this paper, we have analyzed the rates of instability of two types (single- and double-) of periodic waves for a fourth-order NLS equation. We have divided the method of computing instability rate of the single-periodic waves ($dn$- and $cn$-) into two parts. In the first part, we have obtained the eigenvalues and periodic wave solutions by solving the Lax pair equations analytically. We have also found the values of certain undetermined parameters that exist in the considered solution. In the second part, we have introduced a linear perturbation term in the periodic wave solutions and derived a system of linearized equations  whose solutions describe the stability/instability of these waves. By relating the variables of these equations to another spectral parameter $\Lambda$ we have studied the instability of the solutions. We have also investigated the instability of the waves, which has periodicity both in space and time for the considered equation. By considering the solutions in the periodic form and invoking Floquet theory, we have obtained the Lax spectrum (fixed $t$) and stability spectrum (fixed $x$) of these periodic waves. We therefore demonstrated the instability of double-periodic solutions through this simple algorithm and estimated the rate of instability of it for two different values of $\gamma$. It has been shown that, the rate of instability of the double-periodic waves for the NLS equation is smallar when compare to their single-periodic waves \cite{pelinovsky}. In this paper, our investigations revealed that the instability of the double-periodic waves is larger due to the fourth-order dispersion parameter $\gamma$ than that of single-periodic waves. Further, we have also observed that the instability rate increases when the value of system parameter $\gamma$ is increased. According to Ref. \cite {yin}, the properties of MI can affect the width of the optical RWs, and our results would be useful to those studies. Through this study, we have outlined the properties of MI gain for two different periodic wave backgrounds for a fourth-order NLS Eq. (\ref{s21}).
	
	\section*{Acknowledgments}
		NS wishes to thank MoE-RUSA $2.0$ Physical Sciences, Government of India for sponsoring a Fellowship to carry out this work. SR acknowledges MoE-RUSA $2.0$ Physical Sciences, Government of India for providing financial support to achieve the research aim. MS acknowledges MoE-RUSA $2.0$ Physical Sciences, Government of India for sponsoring this research work.
	
	\section*{Authors Contributions}
	All the authors contributed equally to the preparation of this manuscript.
	\section*{Data Availability Statement}
	The data that support the findings of this study are available within the article.
	
\end{document}